\documentclass{aastex62}

\graphicspath{{../code/figs/}}
\usepackage{comment}
\usepackage{amsmath,amssymb}
\usepackage{bm}
\usepackage{enumitem}

\renewcommand{\d}{\mathrm{d}}

\newcommand{\like}{\mathcal{L}}
\newcommand{\given}{\,|\,}
\newcommand{\prior}{\mathcal{P}}

\begin{document}

\title{
  On the Inference of a Star's Inclination Angle from its Rotation
    Velocity and Projected Rotation Velocity
  }

\correspondingauthor{Kento Masuda}
\email{kmasuda@ias.edu}

\author[0000-0003-1298-9699]{Kento Masuda}
\affil{Institute for Advanced Study, 1 Einstein Drive, Princeton, NJ
  08540, USA}

\author[0000-0002-4265-047X]{Joshua N.~Winn}
\affil{Department of Astrophysical Sciences, Princeton University,
Princeton, NJ 08544, USA}

\begin{abstract}

  It is possible to learn about the orientation of a star's rotation
  axis by combining measurements of the star's rotation velocity ($v$)
  and its projection onto our line of sight ($v\sin i$).  This idea
  has found many applications, including the investigation of the
  obliquities of stars with transiting planets.  Here, we present a
  method for the probabilistic inference of the inclination of the
  star's rotation axis based on independent data sets that constrain
  $v$ and $v\sin i$.  We also correct several errors and
  misconceptions that appear in the literature.

\end{abstract}


\section{Introduction}

When studying a single unresolved star, the inclination of the star's
spin axis relative to the line of sight is often irrelevant.  There
are exceptions, though, in which the star's orientation has a
practical or physical significance.  From a practical point of view,
constraints on the stellar inclination can help to interpret
interferometric observations of the stellar disk \citep[see,
  e.g.,][]{Monnier+2012}.  Knowledge of the inclination is also useful
for interpreting photometric or spectroscopic variability due to
starspots, by breaking some of the usual modeling degeneracies
\citep{Walkowicz+2013}.  From a physical point of view, measuring the
inclination angles of large samples of stars can be used to test for
large-scale correlations in spin orientation
\citep{Struve1945,Abt2001,Corsaro+2017,Kuszlewicz+2019}.

The inclination of a star is also of direct physical
significance when it has a disk or an orbiting companion.  In such
cases, the star's spin orientation is a fundamental geometric property that may
relate to the formation and evolution of the system.  Investigations
have been undertaken to assess the angle between the stellar
rotation axis and the orbital and spin axis of a stellar companion
\citep{Hale1994, Albrecht+2009}, the plane of a surrounding disk
\citep{2011MNRAS.413L..71W,2014MNRAS.438L..31G}, and the orbital plane
of a planetary companion \citep{2010ApJ...719..602S}.  The application
to planets has been especially productive because of the large sample
of transiting planets that has recently become available.  The orbit
of a transiting planet always has an inclination close to $90^\circ$.
Therefore, any constraint on the stellar inclination is also a
constraint on the stellar obliquity.

A commonly used method for probing the stellar inclination angle is to
combine measurements or estimates of three quantities: the stellar
radius $R$, the stellar rotation period $P$, and the
line-of-sight projection of the stellar rotation velocity $v\sin i$.
The radius might be based on the observed spectroscopic
parameters, the Stefan-Boltzmann law, or eclipse observations.  The
rotation period comes from the observation of periodic photometric
variability, or from empirical relations between rotation and activity
indicators.  The projected rotation
velocity can be determined from the Doppler-rotational broadening of
the star's absorption lines.  Neglecting the effects of differential
rotation, the inclination can be calculated as
\begin{equation}
\label{eqn:incl}
i = \sin^{-1}\left( \frac{v\sin i}{v} \right) = \sin^{-1} \left( \frac{v\sin i}{2\pi R/P} \right).
\end{equation}
This idea dates back at least as far as
\citet{Abt+1972}, \citet{Doyle+1984}, and \citet{CampbellGarrison1985}.  The latter authors had in mind the Doppler
technique for exoplanet detection, which does not permit the
measurement of a planet's mass $m$, but rather 
$m\sin i$.  But if one is willing to assume that the planet's orbital
axis and the star's spin axis are aligned, then any constraint on
the stellar inclination can be used in conjunction with Doppler data
to determine the planet's mass, or place an upper limit stringent enough
to confirm that the unseen companion is a planet.  This was part
of the evidence presented by \citet{1995Natur.378..355M} that
51~Peg~b is a planet rather than a brown dwarf, although we now know
that spin-orbit alignment is not guaranteed
\citep[see, e.g.,][]{2015ARA&A..53..409W, Triaud2018}.

The subject of this paper is the probabilistic inference of the
stellar inclination angle for the case in which the measurement
uncertainties cannot be neglected and may have nontrivial
distributions.  This is a common situation when dealing with Sun-like
stars, for which $v\sin i$ is only a few kilometers per second, and
the effects of rotation on the stellar line profiles are comparable to
those of turbulence and instrumental broadening.  Furthermore, for any
type of star, the inference of the stellar radius can lead to
asymmetric uncertainty intervals because of the nonlinear
relationships between the observed spectroscopic parameters and the
outputs of theoretical stellar-evolutionary models.  The rotation
period can be determined precisely from a periodic pattern of
photometric variations. However, it is subject to large uncertainties
if only a portion of a rotation cycle is observed, or if the rotation
period must be estimated less directly via activity indicators.

The statistical inference of inclination angles has been the goal of
many studies, including those cited earlier in this paper.  We have
come to realize that many of the methods described in the literature
suffer from errors.  In fact, we have not found any cases that
directly dealt with the probability distribution for $i$ in which the
analysis was 100\% correct.  Some investigators performed simple
``quadrature sum'' error propagation with reference to
Equation~\ref{eqn:incl} assuming independent Gaussian uncertainties
for $v\sin i$, $R$, and $P$.  This is not correct when the
uncertainties are large.  It is also wrong because $v\sin i$ and
$v\equiv 2\pi R/P$ are not statistically independent, i.e., knowledge
of $v$ provides some information about $v\sin i$, and vice versa.
Many investigators who have performed Bayesian analyses have made the
mistaken assumption that $v\sin i$ and $v$ are statistically
independent. In this paper, we present a mathematically correct
procedure based on reasonable assumptions, and explain why some of the
procedures described in the literature are incorrect.

\section{Assumptions}\label{sec:assumptions}

We consider a situation in which we have two data sets, $d_v$ and
$d_u$, from which we have computed the likelihood functions for $v$
and $u \equiv v\sin i$:
\begin{equation}
\like_v(v) \equiv p(d_v\given v), \quad \like_u(u) \equiv p(d_u\given u).
\end{equation}

In practice, the constraint on $v$ usually comes from multiple data
sets.  For example, a photometric time series from which the rotation
period can be measured, and the parallax and spectral energy
distribution from which the star's radius can be determined. These
data can be fitted simultaneously to obtain the likelihood function
for $v=2\pi R/P$.\footnote{The identification of $v$ with $2\pi R/P$
  ignores the effects of latitudinal differential rotation, which
  would otherwise require us to distinguish between the equatorial
  rotation period and the rotation period of whichever features are
  producing the photometric variability.} For simplicity, we denote
the combined data set by $d_v$, and assume that all the relevant
information is incorporated into the likelihood function $\like_v(v)$.

Likewise, we denote by $d_u$ the data set that constrains $u$. The
type of data we have in mind is a high-resolution spectrum that shows
the effects of Doppler-rotational broadening on the star's spectral
lines.  The amount of broadening depends only on the projected
rotation rate and not the sense of rotation, i.e., it does not matter
whether we see the star's north pole or south pole.  Consequently, the
data do not allow $i$ and $180^\circ-i$ to be distinguished.  For this
reason, we restrict our attention to values of $i$ within the range
from $0^\circ$ to $90^\circ$.  The values of $\sin i$ and $\cos i$ are
also between 0 and 1.

Our goal is to combine the information on $v$ and $u$ to compute the
marginal likelihood and posterior probability density function (PDF)
for $\cos i$.  We deal with $\cos i$ rather than $i$ or $\sin i$
because the PDF for $\cos i$ reduces to a constant when the
orientation of the spin axis is completely random.

We also make the following assumptions:
\begin{enumerate}[label=\Roman*.]
\item The data sets $d_v$ and $d_u$ are independent. By this, we
    mean that if the totality of data is $D= \{d_v, d_u\}$, then the
    likelihood function for the $D$ is separable:
\begin{equation}
\label{eq:correct_L_full}
\like_{vu}(v,u) = p(D\,|\,v,u) =
p(d_v|v, u)\,p(d_u|v, u)=p(d_v|v)\,p(d_u|u)=\like_v(v)\,\like_u(u).
\end{equation}

\item The quantities $v$ and $i$ are {\it a priori} independent, i.e.,
  the prior PDF $\prior_{vi}(v,i)$ for these parameters is separable:
\begin{equation}
	\prior_{vi}(v, i) = \prior_v(v)\,\prior_i(i),
\end{equation}
or equivalently,
\begin{equation}
  p(v\given i)=\prior_v(v)~{\rm and}~
  p(i\given v)=\prior_i(i).
\end{equation}
\end{enumerate}

Assumption II is not trivial, and there are circumstances in which it
would be false. For example, if there are populations of stars within
which the stars tend to be spin-aligned, and the populations have
systematically different rotation velocities (because of differences
in age), then there would exist an overall correlation between $v$ and
$i$.  A correlation might also arise from observational bias: a sample
of stars with measured rotation periods may be biased against
low-inclination systems, because the amplitude of the photometric
variability associated with rotation tends toward zero as the star
approaches a pole-on orientation.  As a result, in a signal-to-noise
limited sample, the stars with the weakest rotational variability —--
which tend to be the slower rotators —-- might preferentially have
high inclinations.  For this paper, though, we neglect any such
effects.

Even under Assumption II, we note that $v$ and $u$ are in general
prior dependent: $\prior_{vu}(v,u) \neq \prior_v(v)\,\prior_u(u)$ and
$p(u\given v) \neq \prior_u(u)$.  Ignoring this dependence turns out
to be the main source of errors in the literature. As a demonstration
that $u$ and $v$ are often prior dependent, let us consider a
mathematically simple case in which $\prior_v$ is uniform between $0$
and $v_{\rm max}$, and all possible orientations for the spin axis are
equally probable:
\begin{equation}
	\label{eq:i_prior}
	\prior_{\cos i}(\cos i) = 1, \quad
	\prior_{\sin i}(\sin i) = \frac{\sin i}{\sqrt{1-\sin^2 i}}.
\end{equation}
First,
\begin{equation}
  p(u \given v) = p(v\sin i\given v) = \frac{1}{v}\,p(\sin i \given v) =
  \frac{1}{v}\,\prior_{\sin i}(\sin i) =
  \frac{1}{v}\,\frac{u/v}{\sqrt{1-(u/v)^2}},
\end{equation}
where the third equality follows from Assumption II.
The function 
$p(u\given v)$
rises monotonically as $u$ approaches $v$.
This is a restatement of the familiar fact that low inclinations are
less probable than high inclinations for random spin-axis
orientations. On the other hand,
the prior PDF for $u=v\sin i$ is
\begin{equation}
	\prior_{u}(u)
	=\int_{u \over v_{\rm max}}^1 {1\over s}\, \prior_v\left({u\over s}\right)\,\prior_{\sin i}(s)\,\mathrm{d}s
	= {1\over v_{\rm max}}\int_{u \over v_{\rm max}}^1 {\mathrm{d}s \over \sqrt{1-s^2}}
	= {1\over v_{\rm max}}\left[{\pi \over 2} - \arcsin\left(u \over v_{\rm max}\right)\right],
\end{equation}
where the first equality relies on Assumption II.
This is not the same as  $p(u\given v)$.
Rather, $\prior_u(u)$ is nearly constant for $u \ll v_{\rm max}$, and decreases
as $u$ approaches $v_{\rm max}$.

\section{Correct Procedure}

From the likelihood function for the whole data set, $p(D|v, \cos i)$, we
calculate the marginal likelihood for $\cos i$:
\begin{equation}
	\label{eq:correct_L}
	p(D\given \cos i)
	=\int p(D\given v, \cos i)\,p(v\given \cos i)\,\d v
	=\int \like_v(v)\,\like_u(u)\,p(v\given \cos i)\,\d v
	=\int \like_v(v)\,\like_u(v\sqrt{1-\cos^2 i})\,\prior_v(v)\,\d v.
\end{equation}
Here, the second equality relies on Assumption I,
and the last equality relies on Assumption II
that $p(v\given \cos i)$ is equal to the prior PDF for $v$.
We can obtain the (unnormalized) posterior PDF
for $\cos i$ using Bayes' theorem:
\begin{equation}
	\label{eq:correct-v}
	p(\cos i\given D) \propto
        \prior_{\cos i}(\cos i)\int \like_v(v)\,\like_u\left(v\sqrt{1-\cos^2 i}\right)\,\prior_v(v)\,\d v.
\end{equation}
Using $v=u/\sqrt{1-\cos^2 i}$,
we may also write this as an integral over $u$:
\begin{equation}
	\label{eq:correct-u}
	p(\cos i|D) \propto \frac{\prior_{\cos i}(\cos i)}{\sqrt{1-\cos^2 i}}
          \int \like_v\left({u \over \sqrt{1-\cos^2 i}}\right)\,\like_u(u)\,\prior_v\left({u \over \sqrt{1-\cos^2 i}}\right)\,\d u.
\end{equation}

Given uninformative measurements of $v$ and $v\sin i$, these formulas
should reduce to the prior PDF for $\cos i$.  Indeed, if $\like_u(u)$
is a constant for the values of $u=v\sqrt{1-\cos^2 i}\leq v$ for which
both $\like_v(v)$ and $\prior_v(v)$ have non-zero values, the integral
in Equation~\ref{eq:correct-v} does not depend on $\cos i$ and
therefore $p(\cos i\given D)\propto \prior_{\cos i}(\cos i)$.

\section{Incorrect Procedures}\label{sec:incorrect}

Here we describe some incorrect procedures that have appeared in the
literature, and explain why they are incorrect. In short, these errors
stem from not taking into account the dependence of different
variables.

\subsection{Incorrect Monte Carlo Sampling}
\label{sec:incorrect-sampling}

It is {\it incorrect} to calculate $p(\cos i|D)$ by sampling $v$ and
$v\sin i$ independently from their respective PDFs and constructing
the resulting distribution of $\cos i=\sqrt{1-(v\sin i/v)^2}$.  This
is because $v$ and $v\sin i$ are not statistically independent, as
noted in Section \ref{sec:assumptions}.  For example, it is always the
case that $v\sin i\leq v$. Also, if we assume the ``isotropic priors''
represented in Equation~\ref{eq:i_prior}, then $v\sin i$ is more
likely to be closer to $v$ than to $0$.  Independently sampling from
the PDFs of $v$ and $v\sin i$ does not take into account these
correlations.

This type of mistake has been made frequently \citep[see, e.g.,][]
{HenryWinn2008}.  Two
particular studies deserve mention.  \cite{2014ApJ...796...47M} noted
that this procedure is incorrect but did not supply the reason, and
then proceeded to use a different incorrect method (see Section
\ref{sec:incorrect-marginalization}).  Likewise,
\citet{2014ApJ...783....9H} used a correct formula for $p(i\given D)$
in one part of their study to display the PDFs of $i$ (their Equation
10), but then went on to perform an incorrect Monte Carlo sampling
(their Section 4.3).

\subsection{Incorrect Use of the Product Rule}
\label{sec:incorrect-product}

It is {\it incorrect} to compute the PDF of $\sin i=v\sin i/v$ applying the well-known product rule for $z=xy$:
\begin{equation}
        \label{eq:product}
	p_z(z)=\int p_x(x)\,p_y\left(z\over x\right)\,{1 \over |x|}\,{\d x},
\end{equation}
where $p_x$ and $p_y$ are the PDFs of $x$ and $y$. This is because the
preceding formula is derived under the assumption that $x$ and $y$ are
statistically independent variables, which is not the case for $v$ and
$u$ (Section \ref{sec:assumptions}).  When they are dependent, the
correct formula is
\begin{equation}
	\label{eq:product_correct}
	p_z(z)=\int p_x(x)\,p_y\left(\left. {z\over x}\,\right|\,x\right)\,{1\over |x|}\,\d x.
\end{equation}
For $x=v^{-1}$ and $y=v\sin i$, Equation~\ref{eq:product_correct} yields
\begin{equation}
  p(\sin i \given D) =
  \int p(v^{-1}|D)\,p(\sin i/v^{-1}|v^{-1}, D)\,{1\over v^{-1}}\,\d v^{-1} =
  \int p(\sin i|v, D)\,p(v|D)\,{\d v},
\end{equation}
which is simply the marginal PDF for $\sin i$.  Application of Bayes'
theorem gives
\begin{align}
\notag
p(\sin i \given D) &\propto \int p(\sin i\given v, d_v, d_u)\,\like_v(v)\,\prior_v(v)\,{\d v}\\
\notag
	      &\propto \int p(d_u\given v, d_v, \sin i)\,\prior_{\sin i}(\sin i)\,\like_v(v)\,\prior_v(v)\,{\d v}\\
	      &\propto \prior_{\sin i}(\sin i)\int \like_v(v)\,\like_u(u)\,\prior_v(v)\,{\d v}.
\end{align}
Thus, we obtain
\begin{align}
  p(\cos i \given D) = p(\sin i|D)\,\left|{\d \sin i \over \d \cos i}\right|
      \propto \prior_{\cos i}(\cos i)\int \like_v(v)\,\like_u(u)\,\prior_v(v)\,{\d v},
\end{align}
which recovers Equation~\ref{eq:correct-v}.

The work by \citet{2018AJ....156..253M} included this type of error in
combination with a second error.  The first error was to confuse the
likelihood function $p(D\given \sin i)$ with the PDF $p(\sin i \given
D)$.  The former is the probability to obtain the data $D$ for a fixed
value of $\sin i$, while the latter is the probability density for
$\sin i$ per unit $\sin i$.  The second error was to apply the product
rule incorrectly, as described in this section, to compute $p(\sin i
\given D)$.  The effects of these two errors partly cancelled each
other, as we will show in Section \ref{sec:examples}.  Had they made
only the single error of applying the incorrect product rule to
calculate $p(\sin i\given D)$, and then correctly transformed the PDF
to obtain $p(\cos i\given D)$, they would have arrived at the
incorrect Monte Carlo result described in the previous section.  This
reflects the fact that using the product rule given by
Equation~\ref{eq:product} is equivalent to ignoring the statistical
dependence of $v$ and $v\sin i$.

\subsection{Incorrect Marginalization}
\label{sec:incorrect-marginalization}

It is correct to write down the joint likelihood function for the entire
data set as
\begin{equation}
	p(D\given u, \cos i)
	=\like_v\left(u \over \sqrt{1-\cos^2 i}\right)\,\like_u(u).
\end{equation}
However, it is {\it incorrect} to derive the marginal likelihood as
\begin{equation}
  \label{eq:incorrect-marginalization}
	\mathrm{(incorrect)} 
	\qquad 
	p(D\given\cos i)=\int p(D\given u, \cos i)\,\prior_u(u)\,\d u=\int \like_v\left(u \over \sqrt{1-\cos^2 i}\right)\,\like_u(u)\,\prior_u(u)\,\d u
\end{equation}
because $u=v\sin i$ and $\cos i$ are not statistically independent as
described in Section~\ref{sec:assumptions}.  This incorrect formula
was used by \citet{2014ApJ...796...47M}.  The correct marginal
likelihood is
\begin{align}
	p(D\given \cos i)&=\int p(D\given u, \cos i)\,p(u\given \cos i)\,\d u
	=\int p(D\given u, \cos i)\,{1\over \sin i}\,\prior_v(v)\,\d u
	=\int \like_v(v)\,\like_u(u)\,\prior_v(v)\,\d v,
	&
\end{align}
which reproduces Equation~\ref{eq:correct_L}.

\section{Examples}\label{sec:examples}

\begin{figure*}[h]
	\gridline{
		\fig{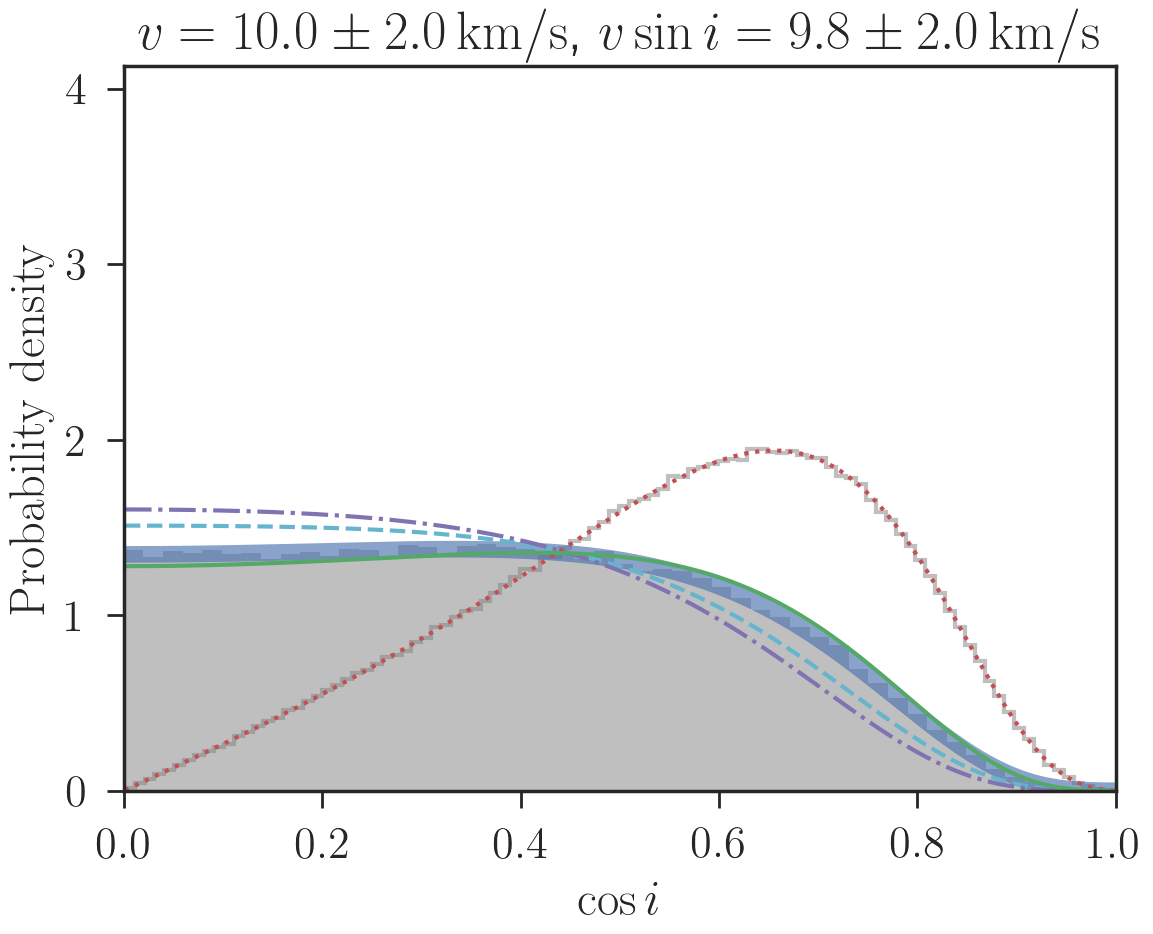}{0.5\textwidth}{}
		\fig{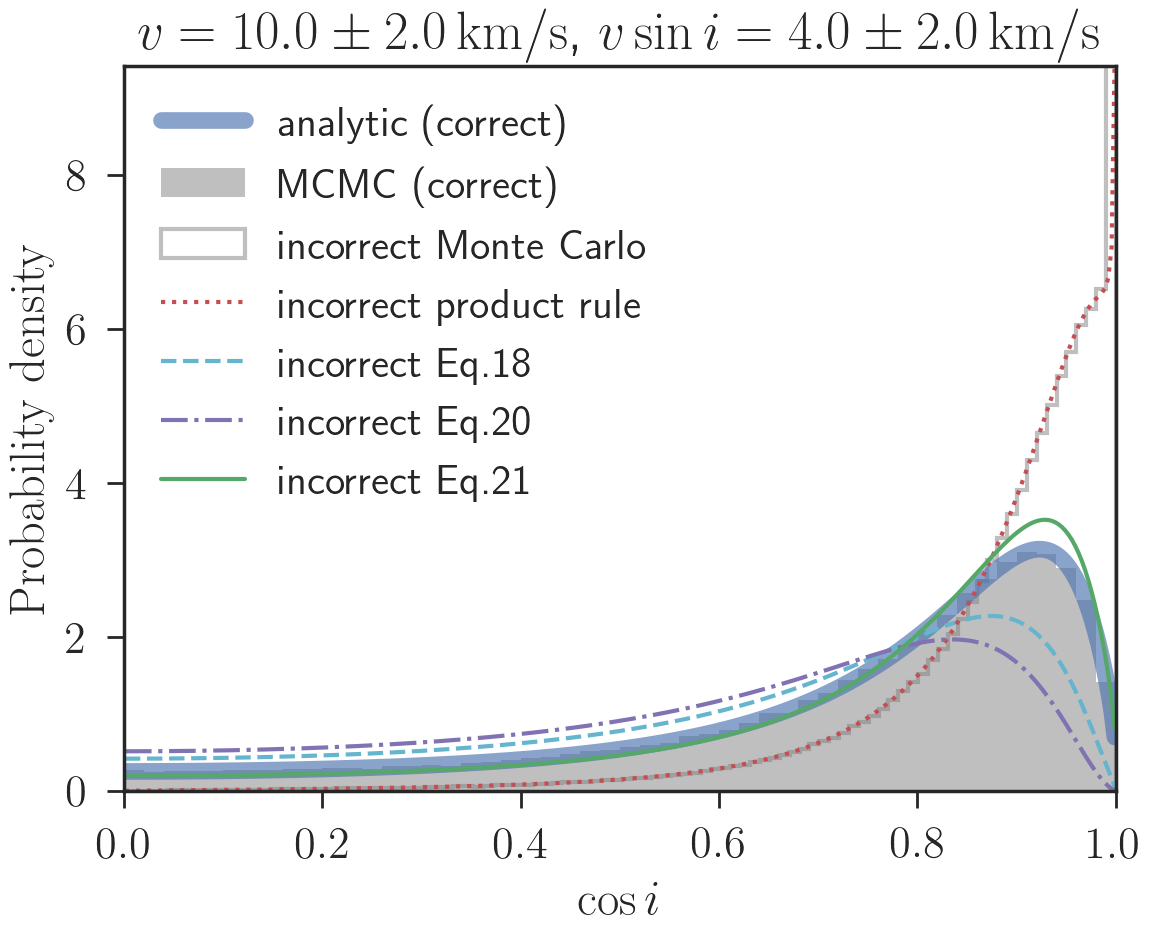}{0.5\textwidth}{}
		}
	\gridline{
		\fig{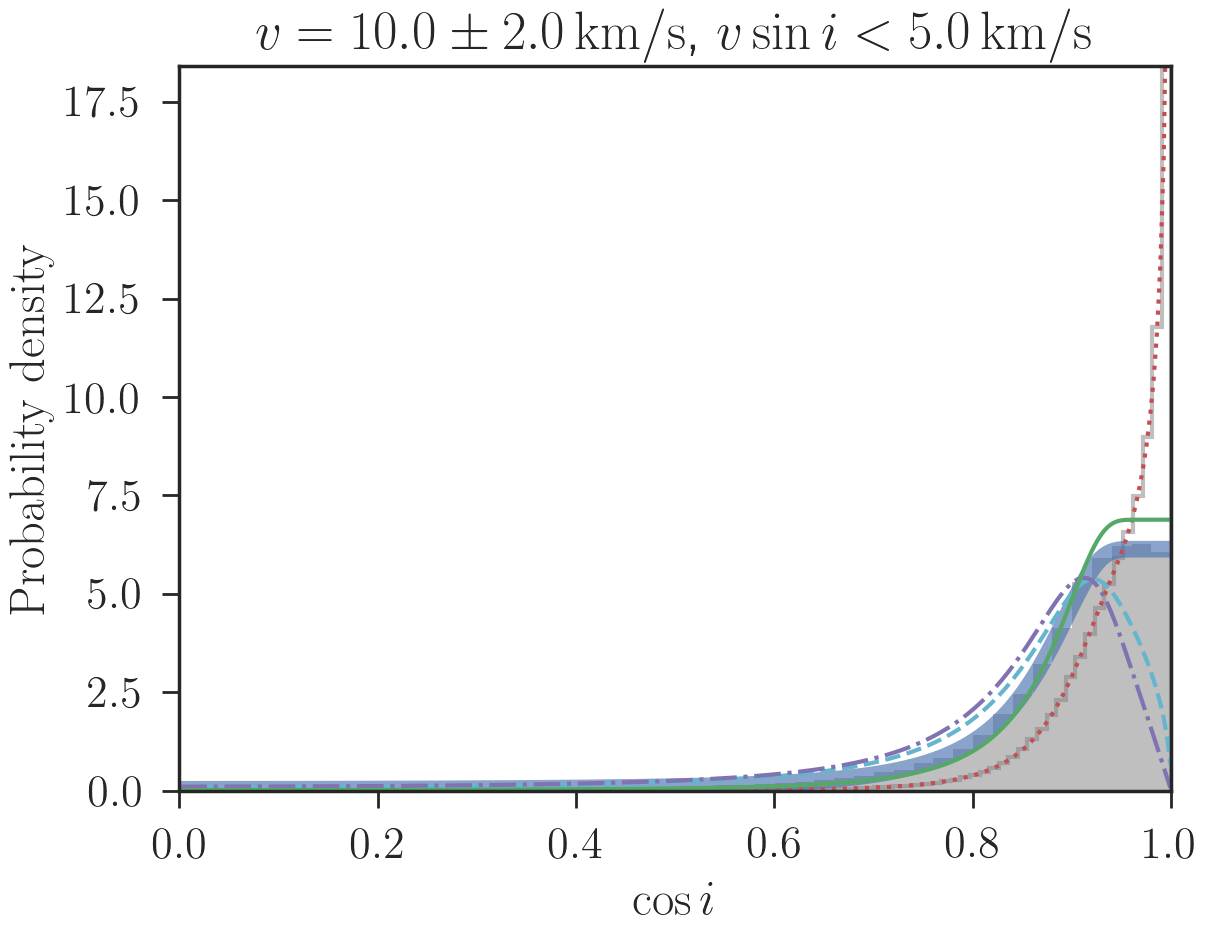}{0.5\textwidth}{}
		\fig{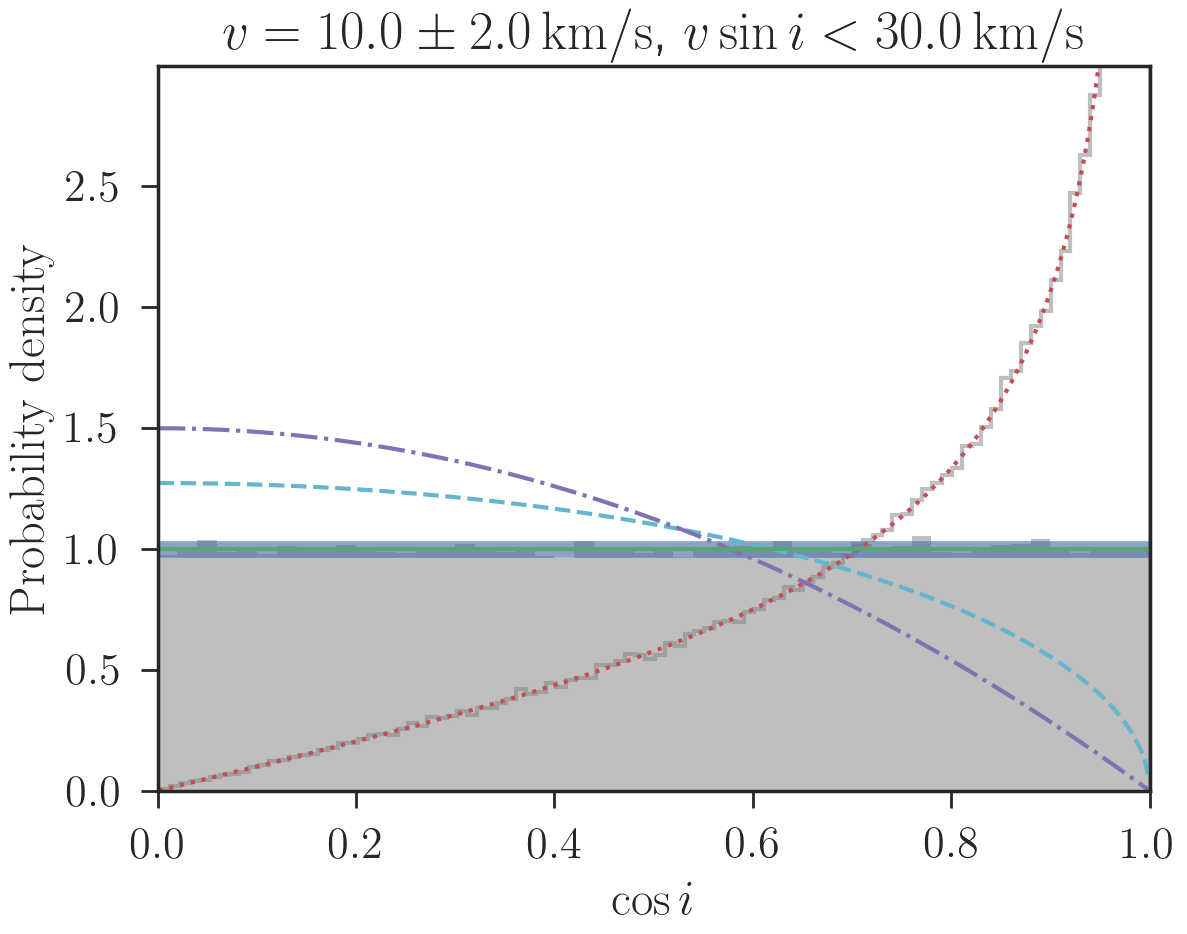}{0.5\textwidth}{}
	}
	\caption{Comparison of different methods to derive the posterior
          probability desnsity function for $\cos i$.}
	\label{fig:comparison}
\end{figure*}

We have calculated $p(\cos i\given D)$ for four illustrative cases,
both by evaluating Equation~\ref{eq:correct-v} directly, and by
performing a Markov Chain Monte Carlo (MCMC) sampling of the posterior
assuming the likelihood given by
Equation~\ref{eq:correct_L_full}.\footnote{We note that the MCMC
  method is not needed to solve this two-dimensional problem: it is
  more efficient to integrate Equation~\ref{eq:correct_L} directly.
  The MCMC method might be worth the computational effort if one deals
  directly with the respective likelihood functions for $v\sin i$,
  $R$, and $P$.}  Throughout this section, we adopt uniform,
normalized, and separable prior PDFs for $v$ and $\cos i$. The four
panels of Figure \ref{fig:comparison} show the results.  The top two
panels show cases in which $\mathcal{L}_v$ and $\mathcal{L}_u$ are
Gaussian functions specified by the central value and dispersion.  The
bottom two panels show cases in which $\mathcal{L}_u$ is non-zero and
flat below a certain threshold value, and zero otherwise. This is
meant to be a mathematically simplified model of a situation in which
only an upper limit on $v\sin i$ is available.  In reality, the
likelihood function $\mathcal{L}_u$ would need to be determined from
the data.

In all of these cases, the distribution of the MCMC samples (filled
gray histograms) agrees with the output of Equation~\ref{eq:correct-v}
(thick blue lines), as expected.  The correct formula recovers the
``uninformative'' PDF for $\cos i$ when the $v\sin i$ measurement has
no constraining power, as shown in the bottom-right panel.  The panels
also show the results of some of the incorrect procedures that have
been used in the literature.  The light gray unfilled histogram is
based on Monte Carlo sampling under the faulty assumption that $v$ and
$v\sin i$ are independent
(Section~\ref{sec:incorrect-sampling}).\footnote{For this simulation,
  we discarded the samples that imply $\sin i \geq 1$.  In some prior
  works, such samples were adjusted to give $\sin i=1$, producing a
  sharp peak at $\cos i=0$.}  The dotted red line is based
on the incorrect application of the product rule for PDFs
(Section~\ref{sec:incorrect-product}), which is equivalent to the
incorrect Monte Carlo sampling.  The dashed blue line was calculated
via incorrect marginalization
(Equation~\ref{eq:incorrect-marginalization}) after multiplying by
$\prior_{\cos i}$ to convert it into a PDF.  The dot-dashed purple
line shows the application of the incorrect formula
\begin{equation}
	\label{eq:incorrect-w17}
  \mathrm{(incorrect)} \qquad 
  p(\cos i | D) \propto \int u\,\like_u(u)
  \like_v\left( \frac{u}{\sqrt{1-\cos^2 i}} \right)\,{\rm d}u,
\end{equation}
based on Equation~4 of \citet{2017AJ....154..270W} after multiplying by
$\prior_{\cos i}$,
and also fixing a typographical error that led to the
subscripts ``1'' and ``2'' being swapped.
The solid green line shows the application of the incorrect formula
\begin{equation}
	\label{eq:incorrect-mp18}
  \mathrm{(incorrect)} \qquad 
  p(\cos i | D) \propto \int v\,\like_u\left( v\sqrt{1-\cos^2 i} \right)
  \like_v(v)\,{\rm d}v,
\end{equation}
from \citet{2018AJ....156..253M} as discussed
in Section \ref{sec:incorrect-product}.

In general, both the incorrect Monte Carlo sampling and the incorrect
application of the product rule lead to a severe overestimation of the
probability density for $\cos i\approx 1$ and an underestimation of
the probability density for $\cos i\approx 0$. Incorrect
marginalization tends to overestimate the probability density at
smaller values of $\cos i$ and significantly underestimate the density
around larger values of $\cos i$. The formula by
\citet{2018AJ....156..253M} is fairly accurate due to the partial
cancellation of errors mentioned previously, but it does tend to
overestimate the probability density for $\cos i\approx 1$.

\section{Summary}

We have described a procedure to combine measurements of $v$ and
$v\sin i$ to derive the marginal likelihood and posterior PDF for
$\cos i$, and have explained why various formulas presented in the
literature are incorrect. The errors originate from the incorrect
treatment for PDFs of the variables that are not statistically
independent from each other. We have also compared the resulting PDFs
from incorrect formulas to the correct PDF for several simplified
models of $v$ and $v\sin i$ measurements.

Based on Figure~\ref{fig:comparison}, we suspect that correcting the
errors made in previous studies using the methods in Section
\ref{sec:incorrect} would not change the qualitative conclusions of
those studies, although it would make quantitative differences.
Regardless of the size of the error, it is always preferred to use the
correct formula if the marginal likelihood or PDF for $i$ is to be
computed analytically.

\acknowledgments We appreciate the referee's constructive comments
which helped to improve the clarity of our presentation.  KM
gratefully acknowledges the support by the W.\,M.~Keck Foundation
Fund.  Work by JNW was partly supported by a NASA Keck PI data award
(contract NNN12AA01C).

\bibliographystyle{aasjournal}

\begin{thebibliography}{}
 \expandafter\ifx\csname natexlab\endcsname\relax\def\natexlab#1{#1}\fi
 \providecommand{\url}[1]{\href{#1}{#1}}
 \providecommand{\dodoi}[1]{doi:~\href{http://doi.org/#1}{\nolinkurl{#1}}}
 \providecommand{\doeprint}[1]{\href{http://ascl.net/#1}{\nolinkurl{http://ascl.net/#1}}}
 \providecommand{\doarXiv}[1]{\href{https://arxiv.org/abs/#1}{\nolinkurl{https://arxiv.org/abs/#1}}}
 
 \bibitem[{{Abt}(2001)}]{Abt2001}
 {Abt}, H.~A. 2001, \aj, 122, 2008, \dodoi{10.1086/323300}

 \bibitem[{{Abt} {et~al.}(1972){Abt}, {Chaffee}, \& {Suffolk}}]{Abt+1972}
 {Abt}, H.~A., {Chaffee}, F.~H., \& {Suffolk}, G. 1972, \apj, 175, 779, \dodoi{10.1086/151598}

 \bibitem[{{Albrecht} {et~al.}(2009){Albrecht}, {Reffert}, {Snellen}, \&
   {Winn}}]{Albrecht+2009}
 {Albrecht}, S., {Reffert}, S., {Snellen}, I. A.~G., \& {Winn}, J.~N. 2009,
   \nat, 461, 373, \dodoi{10.1038/nature08408}
 
 \bibitem[{{Campbell} \& {Garrison}(1985)}]{CampbellGarrison1985}
 {Campbell}, B., \& {Garrison}, R.~F. 1985, \pasp, 97, 180,
   \dodoi{10.1086/131516}
 
 \bibitem[{{Corsaro} {et~al.}(2017){Corsaro}, {Lee}, {Garc{\'\i}a},
   {Hennebelle}, {Mathur}, {Beck}, {Mathis}, {Stello}, \&
   {Bouvier}}]{Corsaro+2017}
 {Corsaro}, E., {Lee}, Y.-N., {Garc{\'\i}a}, R.~A., {et~al.} 2017, Nature
   Astronomy, 1, 0064, \dodoi{10.1038/s41550-017-0064}

 \bibitem[{{Doyle} {et~al.}(1984){Doyle}, {Wilcox}, \& {Lorre}}]{Doyle+1984}
{Doyle}, L.~R., {Wilcox}, T.~J., \& {Lorre}, J.~J. 1984, \apj, 287, 307, \dodoi{10.1086/162689}

 \bibitem[{{Greaves} {et~al.}(2014){Greaves}, {Kennedy}, {Thureau}, {Eiroa},
   {Marshall}, {Maldonado}, {Matthews}, {Olofsson}, {Barlow},
   {Moro-Mart{\'{\i}}n}, {Sibthorpe}, {Absil}, {Ardila}, {Booth},
   {Broekhoven-Fiene}, {Brown}, {Cameron}, {del Burgo}, {Di Francesco},
   {Eisl{\"o}ffel}, {Duch{\^e}ne}, {Ertel}, {Holland}, {Horner}, {Kalas},
   {Kavelaars}, {Lestrade}, {Vican}, {Wilner}, {Wolf}, \&
   {Wyatt}}]{2014MNRAS.438L..31G}
 {Greaves}, J.~S., {Kennedy}, G.~M., {Thureau}, N., {et~al.} 2014, \mnras, 438,
   L31, \dodoi{10.1093/mnrasl/slt153}
 
 \bibitem[{{Hale}(1994)}]{Hale1994}
 {Hale}, A. 1994, \aj, 107, 306, \dodoi{10.1086/116855}
 
 \bibitem[{{Henry} \& {Winn}(2008)}]{HenryWinn2008}
 {Henry}, G.~W., \& {Winn}, J.~N. 2008, \aj, 135, 68,
   \dodoi{10.1088/0004-6256/135/1/68}
 
 \bibitem[{{Hirano} {et~al.}(2014){Hirano}, {Sanchis-Ojeda}, {Takeda}, {Winn},
   {Narita}, \& {Takahashi}}]{2014ApJ...783....9H}
 {Hirano}, T., {Sanchis-Ojeda}, R., {Takeda}, Y., {et~al.} 2014, \apj, 783, 9,
   \dodoi{10.1088/0004-637X/783/1/9}
 
 \bibitem[{{Kuszlewicz} {et~al.}(2019){Kuszlewicz}, {Chaplin}, {North}, {Farr},
   {Bell}, {Davies}, {Campante}, \& {Hekker}}]{Kuszlewicz+2019}
 {Kuszlewicz}, J.~S., {Chaplin}, W.~J., {North}, T. S.~H., {et~al.} 2019,
   \mnras, 488, 572, \dodoi{10.1093/mnras/stz1689}
 
 \bibitem[{{Mayor} \& {Queloz}(1995)}]{1995Natur.378..355M}
 {Mayor}, M., \& {Queloz}, D. 1995, \nat, 378, 355, \dodoi{10.1038/378355a0}
 
 \bibitem[{{Monnier} {et~al.}(2012){Monnier}, {Che}, {Zhao}, {Ekstr{\"o}m},
   {Maestro}, {Aufdenberg}, {Baron}, {Georgy}, {Kraus}, {McAlister}, {Pedretti},
   {Ridgway}, {Sturmann}, {Sturmann}, {ten Brummelaar}, {Thureau}, {Turner}, \&
   {Tuthill}}]{Monnier+2012}
 {Monnier}, J.~D., {Che}, X., {Zhao}, M., {et~al.} 2012, \apjl, 761, L3,
   \dodoi{10.1088/2041-8205/761/1/L3}
 
 \bibitem[{{Morton} \& {Winn}(2014)}]{2014ApJ...796...47M}
 {Morton}, T.~D., \& {Winn}, J.~N. 2014, \apj, 796, 47,
   \dodoi{10.1088/0004-637X/796/1/47}
 
 \bibitem[{{Mu{\~n}oz} \& {Perets}(2018)}]{2018AJ....156..253M}
 {Mu{\~n}oz}, D.~J., \& {Perets}, H.~B. 2018, \aj, 156, 253,
   \dodoi{10.3847/1538-3881/aae7d0}
 
 \bibitem[{{Schlaufman}(2010)}]{2010ApJ...719..602S}
 {Schlaufman}, K.~C. 2010, \apj, 719, 602, \dodoi{10.1088/0004-637X/719/1/602}
 
 \bibitem[{{Struve}(1945)}]{Struve1945}
 {Struve}, O. 1945, Popular Astronomy, 53, 201
 
 \bibitem[{{Triaud}(2018)}]{Triaud2018}
 {Triaud}, A. H.~M.~J. 2018, The Rossiter-McLaughlin Effect in Exoplanet
   Research, in Handbook of Exoplanets (Springer), \dodoi{10.1007/978-3-319-55333-7_2}
 
 \bibitem[{{Walkowicz} {et~al.}(2013){Walkowicz}, {Basri}, \&
   {Valenti}}]{Walkowicz+2013}
 {Walkowicz}, L.~M., {Basri}, G., \& {Valenti}, J.~A. 2013, \apjs, 205, 17,
   \dodoi{10.1088/0067-0049/205/2/17}
 
 \bibitem[{{Watson} {et~al.}(2011){Watson}, {Littlefair}, {Diamond}, {Collier
   Cameron}, {Fitzsimmons}, {Simpson}, {Moulds}, \&
   {Pollacco}}]{2011MNRAS.413L..71W}
 {Watson}, C.~A., {Littlefair}, S.~P., {Diamond}, C., {et~al.} 2011, \mnras,
   413, L71, \dodoi{10.1111/j.1745-3933.2011.01036.x}
 
 \bibitem[{{Winn} \& {Fabrycky}(2015)}]{2015ARA&A..53..409W}
 {Winn}, J.~N., \& {Fabrycky}, D.~C. 2015, \araa, 53, 409,
   \dodoi{10.1146/annurev-astro-082214-122246}
 
 \bibitem[{{Winn} {et~al.}(2017){Winn}, {Petigura}, {Morton}, {Weiss}, {Dai},
   {Schlaufman}, {Howard}, {Isaacson}, {Marcy}, {Justesen}, \&
   {Albrecht}}]{2017AJ....154..270W}
 {Winn}, J.~N., {Petigura}, E.~A., {Morton}, T.~D., {et~al.} 2017, \aj, 154,
   270, \dodoi{10.3847/1538-3881/aa93e3}
 
 \end{thebibliography}

 
\listofchanges
 
\end{document}